\documentclass{ws-ijmpe}
\usepackage{epsfig}

\newcommand{\be}{\begin{eqnarray}}
\newcommand{\ee}{\end{eqnarray}}
\newcommand{\ave}[1]{\left\langle #1 \right\rangle}

\begin{document}

\markboth{Authors' Names}{A statistical model analysis of $K/\pi$
  fluctuations in heavy ion collisions}

\catchline{}{}{}{}{}

\title{A statistical model analysis of $K/\pi$ fluctuations in heavy ion collisions}

\author{Giorgio Torrieri }

\address{Theoretical Physics,
  J.W. Goethe Universitat, Frankfurt A.M., Germany  
torrieri@fias.uni-frankfurt.de}

\maketitle

\begin{history}
\received{22 February 2007}
\end{history}

\begin{abstract}
We briefly describe two statistical hadronization models,
based respectively on the presence and absence of light quark chemical equilibrium,
used to analyze particle yields in heavy ion collisions.
We then try to distinguish between these models using
$K/\pi$ fluctuations data.   We find that while the non-equilibrium model
provides an acceptable description of fluctuations at top SPS and RHIC
energies, both models considerably
under-estimate fluctuations at low SPS energies.   
\end{abstract}

\maketitle
\section{Introduction}  
The exploration of the thermal properties 
of strongly interacting
matter, specifically, of its equation of state, transport coefficients,
degree of equilibration, phase structure, and the dependence of these on the 
energy and system size is one of the main objectives of heavy ion
research.   Thus, it is natural to try to characterize the
soft observables in these collisions using statistical mechanics techniques.

While such an approach has 
a long and illustrious history~\cite{Fer50,Pom51,Lan53,Hag65},      
the systematic and quantitative comparison of data to the statistical hadronization  (SH)
model is   a comparatively recent 
field.
 A consensus has 
developed~\cite{bdm,bdm2,equil_energy,gammaq_energy,jansbook} 
that the SH model can indeed 
fit most, if not all particle yields measured at experiments conducted at
a wide range of energies. Measurements conducted at the GSI Schwerionen Synchrotron
(SIS), BNL's Alternating Gradient Synchrotron (AGS),CERN's Super Proton
Synchrotron (SPS),and BNL's Relativistic
Heavy Ion Collider (RHIC) have successfully been analyzed using SH ansatze.

However, the applied SH models differ regarding
the chemical equilibration condition that is presumed.  
 As a result, it has not as yet been possible to agree statistical
 physics, if any, is responsible for the striking trends observed in the energy
 dependence of some observed hadronic yields.  

In the SH model there are two types of chemical equilibrium:
all models assume relative chemical equilibrium\cite{jansbook}, but some 
also assume absolute chemical equilibrium which implies the presence of   just the right 
abundances of valance  up, down, and strange quark pairs.
If the system of produced hadrons is considered to be in absolute 
chemical equilibrium, then   at highest heavy ion reaction energy  
one obtains chemical freeze-out temperature   $T \sim 160-170$ MeV. 
Values  as low as $T \sim 50$ MeV are reported at lowest reaction energies available.

\begin{figure} 
\begin{center}
\epsfig{width=12cm,clip=,figure=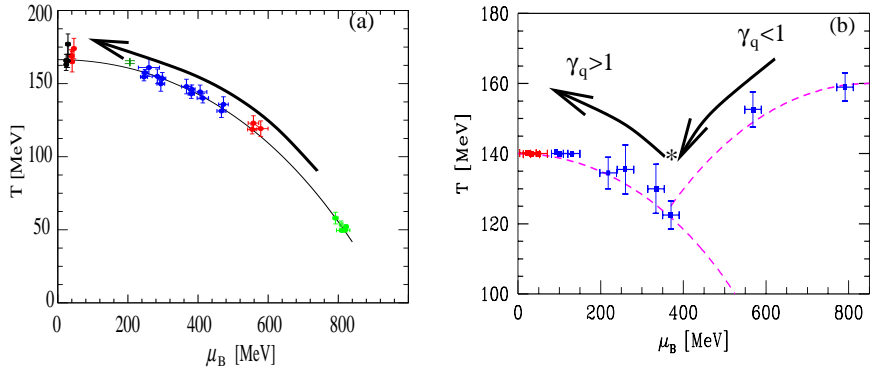}
\vskip -0.5cm
\caption{(Color online)\label{phaseall}
Dependence of freeze-out temperature  $T$ and baryo-chemical potential $\mu_{\rm B}$
on reaction energy in the Equilibrium 
(a), and non-equilibrium (b) 
freeze-out models.  The direction of the arrow corresponds to increasing
$\sqrt{s}$.  The equilibrium dependence of $T$ and $\mu_B$ in the
 panel (a) is not significantly altered by the introduction of the fitted phase
space occupancy $\gamma_s$ and/or the implementation of the Canonical
ensemble for strangeness.
 The ``star'' in panel (b) corresponds to the point where
the transition to the supercooled  regime occurs and the phase space changes
from chemically under-saturated ($\gamma_q<1$) to chemically over-saturated ($\gamma_q>1$).
This point also corresponds to the energy of the ``kink'' and the tip of the ``horn''}
\end{center}
\end{figure}

The energy dependence of the freeze-out temperature than follows the
trend indicated in panel (a) of figure \ref{phaseall}: as the collision energy increases, the freeze-out temperature 
increases and the baryonic density (here baryonic chemical potential $\mu_{\rm B}$)
decreases~\cite{equil_energy}.  An  increase of freeze-out temperature with $\sqrt{s}$ is 
expected on general grounds, since  with increasing reaction energy a 
greater fraction of the energy is carried by mesons created in the collision, 
rather than pre-existing baryons~\cite{Hag65}. 

Further refinements to the equilibrium model, such as the introduction
of the chemical non-equilibrium parameter $\gamma_s$
\cite{jansbook,becattini} and a canonical description of the system
at small energies/system sizes \cite{becattini} do not
materially alter the behaviour of
temperature and chemical potential shown in the panel (a) of
Fig. \ref{phaseall}.

When assuming chemical equilibrium, the variation in the freeze-out
parameters $T$ and $\mu_B$ with energy is remarkably smooth.  However,
there are non-continuous  features in the energy dependence of hadronic observables, 
such as the ``kink'' in the multiplicity per number of participants and the 
``horn''~\cite{horn,horn_theory,gammaq_energy} in certain particle
yield ratios
(top panel of Fig. \ref{figfluct}).   
An effort was made to interpret this in terms of a shift from baryon to meson dominance~\cite{equil_energy}
of the hadron yields. However,  in the chemical equilibrium model even the simple observable 
like $K^+/\pi^+$ remains a smooth function of reaction energy, in contrast to the
experimental results (top panel of Fig. \ref{figfluct}).   
Introduction of $\gamma_s$ and deviations from
the thermodynamic limit, while they help in bringing some of the model
predictions closer to the data, has so far not managed to reproduce
the sharpness of features such as the kink and the horn \cite{becattini}.

Non-monotonic behaviour of particle yield ratios could  indicate
a novel reaction mechanism, e.g. onset of the deconfinement 
phase~\cite{horn_theory}.  In such a situation, the smoothness of the chemical 
freeze-out temperature dependence on energy would be surprising, since it would
imply that  at all energies, from about 1 $A$ GeV at SIS, to the highest RHIC values,
there is no change in either  the fireball evolution dynamics, nor any other  imprint from the 
deconfined phase on the freeze-out condition, which, however is visible in 
the strangeness and entropy yield that $K^+$ and, respectively,
$\pi^+$ represent.   In particular, if the expanding system undergoes a fast conversion
from a Quark Gluon Plasma (QGP) to hadrons,
chemical non-equilibrium and super-cooling~\cite{jansbook} can be
motivated by  entropy and flavor conservation requirements. 

If one abandons the hypothesis of absolute chemical equilibrium, and
fits phase space occupancies for \textit{both} light and strange flavors,
the   systematic behaviour of 
$T$ with energy becomes quite different~\cite{gammaq_energy},
as  is shown  in panel (b) of
figure \ref{phaseall}.   
The two higher $T$ values  at right are for 20 (lowest SPS) and  (most to right) 11.6 $A$ GeV (highest AGS)
reactions. In these two cases the source of particles is  a hot chemically under-saturated  ($T \sim 170$ MeV ) fireball.
Such a system could be a conventional hadron gas fireball that 
had not the time to chemically equilibrate.

Following the thick arrow in panel (b) of
figure \ref{phaseall} we note that  somewhat smaller temperatures are found with further increasing 
heavy ion reaction energies. 
Here it is possible \cite{jansbook}to match the entropy of the emerging 
hadrons with that of a system of nearly massless partons when one considers supercooling
to $T\sim 140$ MeV,  while both light and strange quark phase space in the hadron stage 
acquire significant over-saturation with the phase space occupancy 
$\gamma_{q=u,d}>1$ and at higher energy also $\gamma_s>1$.  A drastic change 
 in the non-equilibrium condition occurs near 30 $A$ GeV,
corresponding to the dip point on right in panel (b) of the figure \ref{phaseall} (marked by an asterisk). 
At heavy ion reaction energy below (i.e. to right in panel (b) of
figure \ref{phaseall}) of this point, hadrons have not reached 
chemical equilibrium, while   at this point, as well as, at heavy ion reaction energy above 
(i.e. at and to left in panel (b) of figure \ref{phaseall}),   hadrons  emerge from a much denser and chemically more saturated system,
as would be expected were QGP formed at and above 30 $A$ GeV.   
This is also  the heavy ion reaction energy corresponding to the ``kink'', which tracks the QGP's entropy density 
(higher w.r.t. a hadron gas), and the peak of the ``horn'' \cite{horn}, 
which tracks the strangeness over entropy ratio (also higher w.r.t. a
hadron gas). 

Distinguishing the two models described here by yields and particle
ratios data has proven to be problematic.  While fits with $\gamma_q$
consistently achieve a higher statistical significance
\cite{gammaq_energy}, the statistical significance of equilibrium
models is well above the conventionally accepted minimum of $5 \%$ at
all energies, so it can not be argued that the non-equilibrium model
is ``objectively better'' at describing data.  Since the two models
are physically different, however, it is also not appropriate
to just assume the equilibrium model holds because of it's greater
``simplicity'' (number of parameters).  Rather, observables should be
found where the two models give different predictions.  In the next
section we will show that fluctuations of $K/\pi$ are an example of such observable.   

\section{$K/\pi$ Fluctuations in chemical (non)equilibrium}  

As shown in \cite{fluctpaper}, fluctuations can be an invaluable test
of light quark chemical non-equilibrium.   This is because the scaled
variance $\sigma_N =\ave{(\Delta N)^2}/\ave{N}$ scales in a different
way from $\ave{N}$ w.r.t. temperature and the light quark phase space
occupancy $\gamma_q$.

An increase in $T$, in general, lowers $\sigma_N$ because of the
greater contribution of resonances that introduce particle
correlations.
An increase in $\gamma_q$, on the other hand, increases $\sigma_N$
where $N$ is a number carried by $\pi$ (such as the multiplicity or
the 
electric charge), since the increased quantum (Bose-Einstein) contributions to
pion yields are enhanced by the phase space over-saturation specific
to $\gamma_q$.

Fluctuations are in general more sensitive than yields to detector
acceptance cuts.  In addition, not understood effects, such as
event-by-event volume fluctuations can give significant contributions
to the observables that are difficult to describe within statistical
models. The solution to second objection, which also resolves part of
the first, is to use ``dynamical'' fluctuations of particle ratios
\begin{equation}
\sigma_{dyn}^{N1/N2} = \sqrt{\sigma_{N1/N2}^2 - (\sigma_{N1/N2}^{stat})^2} 
\end{equation}
where $\sigma_{stat}$ is the fluctuation in mixed events, that, having
no resonance or quantum corrections, should be simply given by
Poissonian scaling. $\sigma_{N1/N2}$ contains such correlation terms,
so
\begin{equation}
\sigma_{N1/N2}^2=\frac{\sigma_{N1}^2}{\ave{N_1}} + \frac{\sigma_{N2}^2}{\ave{N_2}} - 2 \alpha
\frac{\ave{\Delta N_1 \Delta N_2}}{\ave{N_1}\ave{ N_2}} 
\end{equation}
where the correlation term arises due to resonances decaying into
$N_1$ and $N_2$ (and in general has to be weighted by an experimental reconstruction
probability $\alpha$).
We refer the reader to \cite{sharev2} for details
on how to compute the $\sigma$s described here.

\begin{figure} 
\begin{center}
\epsfig{width=10cm,clip=,figure=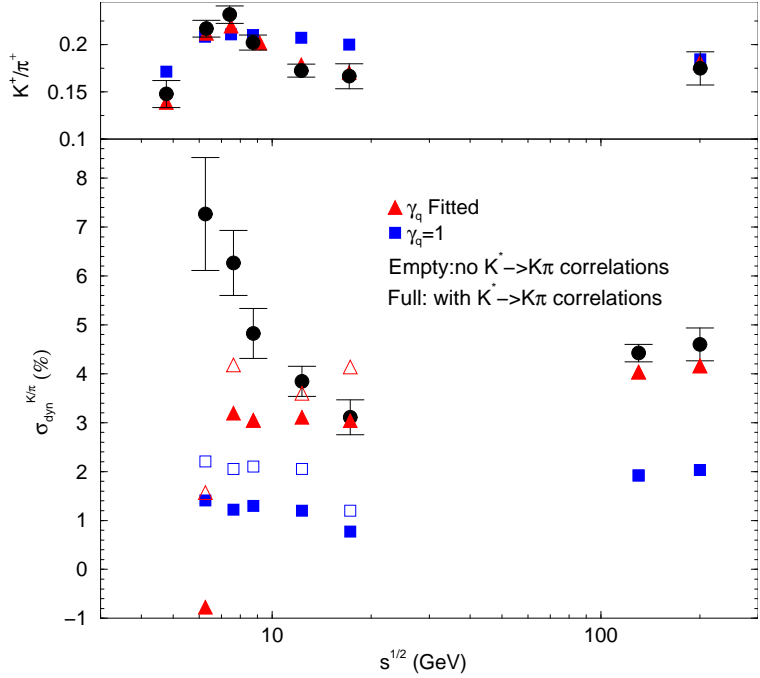}
\caption{\label{figfluct}(Color online)
$K/\pi$ fluctuations with model calculations. Top panels shows the
  ratio dependence on energy}
\end{center}
\end{figure}

We have used the ratios
and yields in the data-samples of \cite{gammaq_energy} to fit
$T,\mu_{B,s}$,$\gamma_s$, and the reaction volume, in the model where
$\gamma_q$ is fitted (non-equilibrium) and where $\gamma_q=1$
(chemical equilibrium).  The results, together with SPS
\cite{spsfluct} and RHIC
\cite{supriya} experimental data  are shown in Fig. \ref{figfluct},
both for the equilibrium model (squares) and the non-equilibrium model
(circles).   We have also shown the two extreme cases of experimental
acceptance, completely preserving the correlations ($\alpha=1$, filled symbols) and
completely destroying them ($\alpha=0$,empty symbols).  The effect of
the correlations is not negligible, but does not
alter the energy dependence of the fluctuations in both models.

As can be seen,$\sigma^{K/\pi}_{dyn}$ at RHIC and the highest SPS
energies is compatible with the non-equilibrium statistical model.
Fluctuations at lower energies, however, can not
be described by any set of chemical parameters that also describes
yields.   As expected from it's smooth
variation of parameters, and smallness of quantum corrections,
$\sigma^{K/\pi}_{dyn}$ does not show a substantial variation in the equilibrium
($\gamma_q=1$ ) model.  This model under-estimates fluctuations at all
energies under consideration.

It should be noted that these results, based on preliminary data,
should be taken with a measure of caution.   The acceptance region of
NA49 is highly non-trivial ($\alpha$ is likely to be neither 0 nor 1,
and strongly momentum-dependant), and our model does not take this into
account.   While the use of dynamical fluctuations eliminates a
\textit{some} of the dependence of the results acceptance cuts, the
acceptance dependence on \textit{correlations} (for instance, $K^*
\rightarrow K \pi$) remains, and introduces a systematic error in our
model.   For RHIC results, $\sigma_{K^{\pm}/\pi^{\pm}}$, this
correlation is negligible since lower-lying resonances do not produce
like-sign particles in their decays.
For SPS results, $\sigma_{(K^{+}+K^{-})/(\pi^+ + \pi^-)}$, the
correlation term is non-negligible but not dominant ($\sim K^*/K \sim
10\%$), as can also be seen from the compatibility of top energy SPS
and RHIC fluctuations.

For $p/\pi$ fluctuations the greater contribution of resonance correlations
($\Delta,\Lambda,\Sigma,\Sigma^*$ and $\Xi$s all decay into $p \pi$,
and form $\sim 70\%$ of the proton yields at all $\sqrt{s}$),
 makes the statistical
description of these fluctuations much less reliable without the
inclusion of acceptance cuts.    For this reason, we do not use this
fluctuation in our analysis, beyond noting that \textit{both} the equilibrium and
non-equilibrium models severely \textit{over}-estimate it
\cite{hq2006}.    We await the measurement of this fluctuation at
RHIC, whose detectors have a very different acceptance region than
SPS's NA49, to fully ascertain the role of limited acceptance in this
fluctuation's measured value.

We can only speculate about the reason for why fluctuations at low SPS
energies are so above the statistical model estimate.
Initial conditions have now been found, at RHIC, to be highly inhomogeneous.  \textit{ if} the system produced in heavy ion collisions is
perfectly thermalized throughout it's evolution (``a perfect fluid''), than the only trace these
inhomogeneities leave at freeze-out is a fluctuation in the system
volume, that is cancelled event by event when fluctuations of ratios are
considered.   

In a system that is \textit{not} perfectly equilibrated, however,
``kinetic'' fluctuations due to the random nature of each collision
between the system's degrees of freedom can indeed arise.   Perhaps
this is the origin of the larger $K/\pi$ fluctuations observed at
lower SPS energies, either due to a greater impact of the hadronic
``corona'', or lack of equilibration in the whole system.

To test this speculation, it is necessary to model $K/\pi$
fluctuations through kinetic models that include the fluctuation
in the initial condition, ``dynamical'' fluctuations due to the finite
number of dynamical processes that generate Kaons and $\pi$s (as well
as the microscopic randomness of each process) and a model of the detector's acceptance cuts
(as mentioned, highly non-trivial for NA49).
Such a study goes beyond the scope of this work.

In conclusion, we have given a description of the two statistical
models currently on the market.  As a way to differentiate between these models,
we have calculated the $\sigma^{dyn}_{K/\pi}$ as a function of energy,
and compared the model predictions to experimental data.
The results, while interesting, are not conclusive:  The equilibrium
model under-estimates $\sigma^{K/\pi}_{dyn}$ at all energies, while the non-equilibrium model describes the higher
energies SPS and RHIC acceptably.  It however considerably
under-estimates $\sigma^{K/\pi}_{dyn}$ at
lower SPS energies.   While this might be an indication that the
system at these energies is not completely thermalized, and hence the
random nature of each microscopic interaction enhances the final
fluctuations beyond the statistical expectation, this can not for now
be conclusively established.

The author thanks the Von Humboldt foundation for the financial
support given, Frankfurt University for the hospitality provided,
as well as  the QM2006 conference organizers for
the financial support that allowed him to attend the conference.   We also thank Michael Hauer, Mark
Gorenstein, Marek Gazdzicki, Marcus Bleicher, Volker Koch, Sangyong Jeon, and Johann Rafelski for stimulating discussions.

\end{document}